\def\bbo{{\mbox{\bf b}}}
\def\rbo{{\mbox{\bf r}}}
\def\xbo{{\mbox{\bf x}}}
\def\ybo{{\mbox{\bf y}}}

\def\qbar{\bar{q}}

\newcommand{\be}{\begin{equation}}
\newcommand{\ee}{\end{equation}}
\newcommand{\beeq}{\begin{eqnarray}}
\newcommand{\eeeq}{\end{eqnarray}}

\documentclass[10pt]{ismd08}
\usepackage{graphicx}
\usepackage{cite,./mcite}

\begin{document}
\title{Theoretical concepts of parton saturation - from HERA to LHC}
\author{Krzysztof Golec-Biernat}
\institute{Institute of Nuclear Physics Polish Academy of 
Sciences, Cracow, Poland,\\ 
Institute of Physics, University of  Rzesz\'ow, Rzesz\'ow, Poland}
\maketitle
\begin{abstract}
We present  a short summary of parton saturation concepts as seen in deep inelastic scattering.
\end{abstract}

\section{Introduction}
\label{sec:1}

The deep inelastic scattering (DIS) experiments, in which leptons probe nucleons with the help
of electroweak bosons, reveal that nucleons consist of partons. These 
are colored quarks of Quantum Chromodynamics (QCD) which carry  approximately  half of the nucleon's momentum. The missing half is provided by gluons to which
the electroweak bosons do not couple. Thus, although not directly probed, gluons
are extremally important for the description of the nucleon structure.
Quantitatively, this is summarized by the DGLAP evolution equations of
QCD which govern the dependence of the quark and gluon
distributions  in a nucleon on a scale $Q^2$ (identified in DIS with photon's virtuality
$q^2=-Q^2$). 
The sign of the logarithmic derivative,
$\partial F_2/\partial \log Q^2$, at different values of the Bjorken variable $x$ is determined
by the relative contribution of quarks to gluons. In the limit $x\to 0$, studied intensively
by the experiments  at HERA, the deep inelastic processes are dominated
by a strongly rising gluon distribution. Therefore, in the small-$x$ limit, 
gluonic systems inside the  nucleon are predominantly studied.
The description of processes in such systems, using perturbative QCD
(pQCD), is the aim of this presentation.

\section{Collinear factorization versus $k_T$-factorization}
\label{sec:2}

\begin{figure}[t]
     \centerline{\includegraphics[width=6cm]{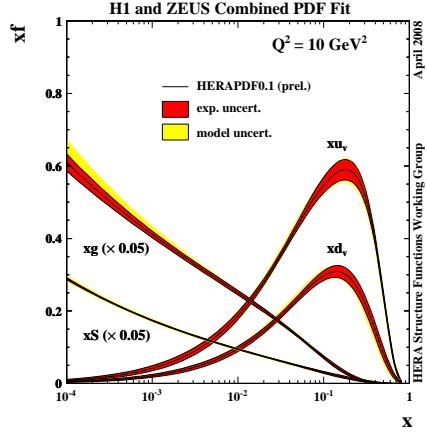}}
\caption{\it Parton distributions from a global fit to the HERA data as functions of $x$ for fixed $Q^2=10~{\rm GeV}^2$.}
\label{fig:1}
\end{figure}

In the electron--proton DIS, the measured proton structure functions, $F_T$ and $F_L$, are  
related to the parton distributions through the collinear factorization formula resulting from
pQCD:
\begin{equation}
\label{eq:1}
F_{T,L}(x,Q^2)=\sum_{i=q,\qbar,g}\{C_{T,L}^{(i)}\otimes f_i)\}(x,Q^2)+\sum_{n=1}\frac{\Lambda_{T,L}^{(n)}(x,\alpha_s)}{Q^{2n}}
\end{equation}
where $\otimes$ indicates integral convolution in parton longitudinal momentum fractions, $\alpha_s=\alpha_s(Q^2)$ 
is the running strong coupling constant, $C_{T,L}^{(i)}(z,\alpha_s)$ 
are perturbatively computed  coefficient functions and $f_i(x,Q^2)$ 
are quark, antiquark and gluon distributions (multiplied by $x$). The $Q^2$-dependence of the 
parton distributions is determind by the DGLAP evolution equations 
\cite{Gribov:1972ri,*Altarelli:1977zs,*Dokshitzer:1977sg} with initial conditions which are fitted to data. The first term on the r.h.s. of eq.~(\ref{eq:1}) provides the leading twist-2 description with logarithmic dependence on $Q^2$ while the remaining terms, called higher twists, seems to be suppresses for large $Q^2$. In the standard analysis, a global fit of the leading twist formula to the HERA data on $F_2=F_T+F_L$, together with cross sections of other hard processes, leads to the determination of the parton distributions shown in Fig.~\ref{fig:1}. A distinct feature of this determination is a strong rise of the gluon and sea quark distributions for $x\to 0$. 

A closer theoretical examination of the small-$x$ scattering reveals that for not too high
$Q^2$, the higher twist
terms cannot be neglected since they are enhanced by powers of $\alpha_s\log(1/x)$,
when the smallness of $\alpha_s$ is compensated by large logarithm of $x$. 
The relevant resummation of such terms in the leading (LO) and next-to-leading (NLO) logarithmic approximation leads to the BFKL approach to the structure functions with the following $k_T$-factorized form
\cite{Catani:1990xk,*Catani:1990eg,*Collins:1991ty}:
\be
\label{eq:2}
F_2(x,Q^2)=Q^2\int \frac{d^2k_T}{k_T^4}\,\Phi(k_T^2/Q^2,\alpha_s(k_T))\,f(x,k_T)
\ee
where the impact factor $\Phi(k_T^2/Q^2,\alpha_s(k_T))$ describes the interaction of
the virtual photon with a gluon with nonzero transverse momentum $k_T$. In the LO this is the
process:~$\gamma^*(Q^2)g(k_T)\to q\qbar$.
The function
$f(x,k_T)$ is called unintegrated gluon distribution which obeys the BFKL equation
\cite{Fadin:1975cb,*Lipatov:1976zz,*Kuraev:1977fs,*Balitsky:1978ic} and is
related  to the gluon distribution $g(x,Q^2)$ through the formula
\be\label{eq:3}
xg(x,Q^2)=\int \frac{d^2k_T}{k_T^2}\,f(x,k_T)\,\theta(|k_T|<Q^2)\,.
\ee 
From the solution of the BFKL equation, the small-$x$ limit is dominated by 
the gluon  distribution with the power-like rise, $f(x,k_T)\sim x^{-\lambda}$  and $\lambda\approx 0.3$. There is a general agreement, based on the experience with the Froissart-Martin bound, that such a rise of the gluon distribution, and in consequence $F_2$, 
violates unitarity and eventually
must be tamed. The BFKL solution is also plagued by diffusion to infrared, namely, the $k_T$-integration in the pQCD formula (\ref{eq:2}) is quickly  dominated by the contribution from the soft momenta region, $k_T\approx \Lambda_{QCD}$, where the Landau pole of $\alpha_s(k_T)$ is encountered.
A cure for these problems is absolutely necessary.
 
\section{Parton saturation}
\label{sec:3}

\begin{figure}[t]
     \centerline{\includegraphics[width=5cm]{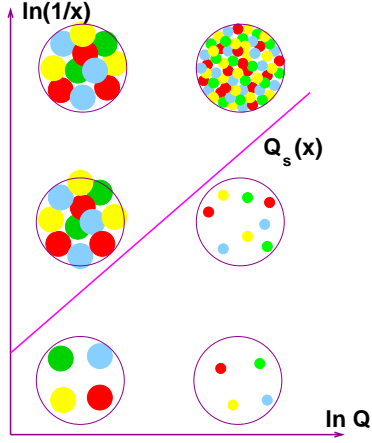}}
\caption{\it Saturation line in the $(x,Q^2)$-plane.}
\label{fig:2}
\end{figure}

The taming of the power-like rise of the gluon distribution $xg(x,Q^2)$ was addressed 
for the first time by Gribov, Levin and Ryskin in \cite{Gribov:1984tu} in the double logarithmic approximation.
Summing fan diagrams,  which take into account the fusion of $t$-channel gluons, the linear DGLAP equation for the gluon distribution receives a negative, nonlinear term,
\be
\label{eq:4}
\frac{\partial^2 xg(x,Q^2)}{\partial \ln(1/x) \partial
\ln Q^2}= \overline{\alpha}_s xg(x,Q^2)-\frac{{\alpha}_s^2}{\pi^2 R^2}
\frac{[xg(x,Q^2)]^2}{Q^2}\,,
\ee
where $\overline{\alpha}_s={N_c\alpha_s}/{\pi}$ and 
the parameter $R$ controls the strength of the nonlinearity.
With such a modification, the gluon distribution saturates for $x\to 0$, and so
does the structure function. This result was extended in \cite{Mueller:1985wy}
by including nonlinear modifications for the sea quark distributions.
A crucial feature introduced by the nonlinearity is an 
$x$-dependent saturation scale $Q_s^2(x)$, defined as a value
of $Q^2$ for which the nonlinear term in eq.~(\ref{eq:4}) is comparable with the linear one:
\be\label{eq:5}
xg(x,Q_s^2)\,\frac{{\alpha}_s(Q_s^2)}{Q_s^2}\,\sim\,\pi R^2\,.
\ee
Therefore, saturation effects are important when
the number of  gluons per unit of rapidity, $xg$, 
times the gluon-gluon interaction cross
section, $\alpha_s/Q^2$, approaches the geometric size of the nucleon or a gluonic system inside the nucleon (``hot spot''). In such a case, a  simple additive treatment of parton emission breaks down and gluons start to annihilate. 
Since from (\ref{eq:5}) $Q_s^2\sim xg$ and 
$xg\sim x^{-\lambda}$ before the saturation limit is reached, we find  
that $Q_s^2\gg \Lambda^2_{QCD}$ for sufficiently small $x$,  and 
the presented approach based on perturbative QCD is justified . 
This is schematically illustrated in Fig.~\ref{fig:2} where two regions separated
by the saturation line, $Q^2=Q_s^2(x)$, are shown. Below this line, in the dilute region,
the linear evolution equations are valid, while approaching the line, the saturation region
is entered with nonlinear equations describing parton saturation.

Eq.~(\ref{eq:4}) is a rather crude approximation since it is  valid in the extreme case, $x\to 0$ and $Q^2\to \infty$. In the $k_T$-factorization approach the latter limit is relaxed
and only large logarithms $\log(1/x)$ are relevant.
Summing  BFKL  pomeron fan diagrams with triple pomeron vertices
in the leading logarithmic approximation and in the limit of large number of colors $N_c$, 
the Balitsky-Kovchegov (BK)  equation for the unintegrated gluon density $\phi(x,k_T)$ is found\footnote{$\phi$ is related to the unintegrated gluon density $f$ from Section \ref{sec:2} by $f(x,k)\sim k^2\nabla_{k}^2\phi(x,k)$.} \cite{Balitsky:1995ub,Kovchegov:1999yj,*Kovchegov:1999ua}:
\be\label{eq:6}
{\partial_Y \phi(x,k_T)}
\,=\,\overline{\alpha}_s\,\chi(-\partial_L)\,\phi\,-\,\overline{\alpha}_s\,
\phi^2
\ee
where  $Y=\log(1/x)$ is rapidity, $L=\log k_T^2$ and $\chi$ is the BFKL characteristic function. 
This nonlinear equation generalizes the linear BFKL equation. The properties of its solutions were intensively studied both  analytically  
\cite{Munier:2003vc,*Munier:2003sj,*Munier:2004xu} 
and numerically \cite{Armesto:2001fa,GolecBiernat:2001if}. 
The most fruitful approach is based on the relation to the known
from statistical physics Fisher-Kolmogorov equation, which admits travelling wave solution.
In our language,
it means that the BK solution develops a saturation scale, $Q_s(x)\sim x^{-\lambda}$ with known value of $\lambda$
\cite{Munier:2003vc,*Munier:2003sj,*Munier:2004xu}, such that for small $x$ we have
\be\label{eq:7}
\phi(x,k_T)=\phi(k_T/Q_s(x))\,.
\ee
This property, called geometric scaling, was observed in the data from HERA 
\cite{Stasto:2000er}. Looking more carefully, for $k_T\gg Q_s(x)$ the gluon distribution 
$\phi\sim 1/k^2_T$, while for  small transverse momenta, $\Lambda_{QCD}\ll k_T< Q_s(x)$, the behaviour changes to logarithmic,
$\phi\sim \ln(Q_s(x)/k_T)$. This is the illustration of the transition 
to saturation, when both the power-like growth in $x$ and infrared  diffusion in $k_T$ 
of the gluon distribution are tamed \cite{GolecBiernat:2001if}, 
see  Fig.~\ref{fig:2} with $Q^2\equiv k_T^2$.

\section{Color dipole approach and beyond}
\label{sec:4}

A more intuitive approach to parton saturation is provided by the color dipole approach \cite{Nikolaev:1990ja,Mueller:1993rr}.
In the target rest frame, the DIS at small $x$ can be formulated as the eikonal scattering of a color quark-antiquark dipole,  formed by the splitting $\gamma^*\to q\qbar$,  
on the target color field. The dipole scattering amplitude
$N(\xbo,\ybo)$ is given by two  Wilson lines collinear to quarks velocity $u$
\be\label{eq:8}
N(\xbo,\ybo)=1-\frac{1}{N_c}{\rm Tr}\,U(\xbo)U^\dagger(\ybo)\,,
~~~~~~
U(\xbo)={\rm P}\exp\left\{ig\int_{-\infty}^\infty d\lambda\, u\cdot A(\lambda u+\xbo)\right\}~
\ee
where $\xbo$ and $\ybo$ are two dimensional vectors of quark transverse positions, conserved
during the collision, and $A$ is a target color field. The deviation of the classcal
quark trajectory from the light-like line defines the change of  $N$ with rapidity $Y$, which
leads to the new BK equation for the dipole scattering amplitude \cite{Balitsky:1995ub}. 
Its solutions fulfil the unitarity bound, $N\le 1$.
When the dependence on the 
impact parameter, $\bbo=(\xbo+\ybo)/2$ is neglected, the new equation is equivalent to eq.~(\ref{eq:6})  after Fourier transforming of $N/\rbo^2$
with respect to $\rbo=\xbo-\ybo$. 
The BK equation in the transverse space was also obtained in the
Mueller's dipole approach \cite{Mueller:1993rr,*Mueller:1994jq,*Mueller:1994gb} in which the $q\qbar$ dipole develops  a system of dipoles (by radiating soft gluons in the large $N_c$
approximation)  which subsequently multiple interact with a large
nucleus target \cite{Kovchegov:1999yj,*Kovchegov:1999ua}.

The dipole scattering amplitude is the basic ingredient in the computation of the nucleon structure functions at small $x$. In the last ten years, this amplitude was also modelled using
the properties of the BK  solutions such as color transparency, $N\sim r^2$ for a small dipole size $r=|\rbo|$; geometric scaling, $N=N(rQ_s(x))$; and the unitarity bound,
$N\le 1$. A recent comprehensive review on the dipole models of DIS processes is presented in \cite{Motyka:2008jk}.

The BK equation describes unitarity corrections in the asymmetric configuration when
the target is extended and dense and the projectile is small and dilute. In a more
symmetric configuration, e.g. in the  $pp$ scattering at LHC, the BK equation is no longer sufficient,
which means that  in the diagrammatic approach
closed pomeron loops have to be taken into account besides  fan diagrams. An interesting attempt in this direction was made
in \cite{Avsar:2005iz,*Avsar:2006jy,*Avsar:2007xg} where pomeron loops were modelled 
as color reconnections in the dipole cascades.
The resulting scattering amplitudes respect the target-projectile symmetry and  describes reasonable well the existing total and diffractive cross sections in the $p\overline{p}$ scattering.
The pomeron loops were also studied in a statistical approach, based on the stochastic
Fisher-Kolmogorov equation, finding a new kind of scaling called diffusive  scaling \cite{Mueller:2004se,*Munier:2005re,*Iancu:2004es}. Recently, high energy factorization
theorems for the gluon production in heavy nucleus collisions were proven in the color
glass condensate approach \cite{Gelis:2008rw,*Gelis:2008ad}.

\medskip
\noindent{\bf Acknowledgements}  This work has been supported  by the Polish grant
no. N N202 249235 and by the Research Training Network HEPTools  (MRTN-2006-CT-035505).
\begin{footnotesize}
\bibliographystyle{ismd08} 
{\raggedright
\bibliography{golec}
}
\end{footnotesize}
\end{document}